\documentclass[%
 reprint,
 amsmath,amssymb,
 aps,
]{revtex4-2}

\usepackage{graphicx}
\usepackage{color}
\usepackage{dcolumn}
\usepackage{bm}
\usepackage{hyperref}
\usepackage{subfigure}
\usepackage{acronym}
\usepackage{CJK}
\usepackage{orcidlink}
\usepackage{braket}

\graphicspath{ {./figures} }

\newacro{GR}{general relativity}
\newacro{GW}{gravitational wave}
\newacro{BH}{black hole}
\newacro{BBH}{binary black hole}
\newacro{QNM}{quasi-normal mode}
\newacro{SNR}{signal-to-noise ratio}
\newacro{IMR}{inspiral-merger-ringdown}
\newacro{JSD}{Jensen-Shannon Divergence}
\newacro{NR}{numerical relativity}

\begin{document}
\begin{CJK*}{UTF8}{gbsn}

\preprint{APS/123-QED}

\title{Contribution from Nonlinear Quasi-normal Modes in GW250114}

\author{Yuxin Yang(杨雨鑫)\orcidlink{0009-0008-9388-0620}$^{1}$}
\author{Changfu Shi(石常富)}\email{shichf6@mail.sysu.edu.cn}
\author{Yi-Ming Hu(胡一鸣)\orcidlink{0000-0002-7869-0174}$^{1}$}\email{huyiming@sysu.edu.cn}
\affiliation{$^1$MOE Key Laboratory of TianQin Mission, %
            TianQin Research Center for Gravitational Physics $\&$ School of Physics and Astronomy, %
            Frontiers Science Center for TianQin, %
            Gravitational Wave Research Center of CNSA, %
            Sun Yat-sen University (Zhuhai Campus), %
            Zhuhai, 519082, China}

\date{\today}

\begin{abstract}
We report evidence for nonlinear gravitational effects in the ringdown signal of gravitational wave event GW250114. Using Bayesian inference, we find that the inclusion of a nonlinear quasi-normal mode (220Q)—a second-order harmonic predicted by general relativity—is statistically favored over the standard linear model (440 mode) when analyzing the post-merger oscillations. Specifically, models incorporating the 220Q mode yield higher Bayes factors than those including only the linear 440 mode, and produce remnant black hole parameters (mass and spin) more consistent with full numerical relativity simulations.
This suggests that nonlinear mode coupling contributes significantly to the ringdown phase, opening a new avenue to probe strong-field gravity beyond linear approximations.
\end{abstract}

\maketitle
\end{CJK*}

\maketitle

\textit{Introduction.---}
Nonlinearity is one of the most prominent features of \ac{GR}.
Since its inception, \ac{GR} has undergone a wealth of experimental tests \cite{Will:2014kxa,Berti:2015itd}, including Solar System tests \cite{Williams:1976zz}, binary-pulsar experiments \cite{Kramer:2021jcw}, observations of massive black holes at galactic centers \cite{GRAVITY:2018ofz,Do:2019txf,EventHorizonTelescope:2019dse}, cosmological measurements \cite{Baker:2014zba,Stairs:2003eg}, and \ac{GW} observations in the past decade \cite{LIGOScientific:2016lio,Abbott:2018lct,LIGOScientific:2019fpa,LIGOScientific:2020tif,LIGOScientific:2021sio}.
These tests cover low-velocity, quasi-static, weak-field regimes, and strong-field with highly dynamic regimes.
The results consistently demonstrate a good agreement between \ac{GR} and experimental data \cite{Will:2014kxa}. However, the detection and verification of the nonlinear effects in \ac{GR} require further investigation \cite{Ioka2007,Okuzumi2008,Shi:2024ttu,Khera:2024bjs}.

The Kerr metric, characterized by two parameters: mass $M$ and spin $\chi$, well describes the \acp{BH} in the Universe~\cite{Kerr:1963ud}.
After a \ac{BBH} merger, as the remnant settles into a stable configuration, the perturbed Kerr \ac{BH} excites the \ac{GW} with a discrete set of \acp{QNM} \cite{Nollert:1999ji,Kokkotas:1999bd,Berti:2009kk,Berti:2005ys}.
The frequencies of \acp{QNM}, known as the \ac{BH} spectrum, can be measured to test whether the inferred mass and spin are consistent with the Kerr hypothesis, thereby offering a stringent test of \ac{GR} \cite{Dreyer2004,Shi2019,Gossan:2011ha}.
\citet{Mitman:2022qdl} and \citet{Cheung:2022rbm} independently observed in \ac{NR} simulations that second-order perturbations of Kerr \ac{BH} trigger quadractic \acp{QNM}.
This makes it possible to test higher-order \ac{GR} effects by detecting such quadratic \acp{QNM} consistent with the Kerr \ac{BH} \cite{Shi:2024ttu,Luo:2025ewp}.

Since the first \ac{GW} event GW150914, hundreds of \ac{BBH} merger events have been observed~\cite{LIGOScientific:2016aoc, LIGOScientific:2018mvr, LIGOScientific:2020ibl, KAGRA:2021vkt, LIGOScientific:2025slb}.
Among those, several of the events with high \acp{SNR} have been analyzed for \acp{QNM} beyond the fundamental one.
In GW190521, convincing evidence was found for the fundamental $(3, 3)$ mode~\cite{Capano:2021etf, LIGOScientific:2020iuh, Siegel:2023lxl}.
The existence of an overtone in GW150914 has long been debated~\cite{Carullo:2019flw, Finch:2022ynt, LIGOScientific:2020tif, CalderonBustillo:2020rmh, Ghosh:2021mrv, Wang:2023ljx, Ma:2023cwe, Ma:2023vvr, Crisostomi:2023tle, Wang:2024yhb, Correia:2023bfn, Isi:2019aib, Isi:2022mhy, Lu:2025mwp, Cotesta:2022pci, Carullo:2023gtf}, but the detection of GW250114 provided strong evidence supporting the overtone $221$ mode~\cite{KAGRA:2025oiz, LIGOScientific:2025obp}, ending the debate whether overtone has ever been detected~\cite{KAGRA:2025oiz, LIGOScientific:2025obp}.
Such loud event also enables a number of interesting science, like revealing the nature of the black hole horizons~\cite{Akyuz:2025seg, Dima:2025tjz, Chandra:2025ipu, Lu:2025vol}.

For this event, the LIGO-Virgo-KAGRA Collaboration claimed the detection of the $440$ mode~\cite{LIGOScientific:2025obp}.
The argument for its presence was made with two independent approaches: the \texttt{pSEOBNR} analysis and the \texttt{KerrPostmerger} model analysis.
The \texttt{pSEOBNR} analysis~\cite{Brito:2018rfr, Pompili:2025cdc, Maggio:2022hre, Ghosh:2021mrv} refers to testing possible deviations in the \acp{QNM} from the \ac{GR} using the \texttt{SEOBNRv5PHM} waveform~\cite{Ramos-Buades:2023ehm, Pompili:2023tna}, which is a full \ac{IMR} waveform calibrated to the \ac{NR} simulations.
When analyzing simulated \ac{NR} data without injecting the $(\ell, |m|) = (4, 4)$ multipoles, the posterior distribution of the frequency deviation of $440$ mode, $\delta \hat{f}_{440}= \frac{\hat{f}_{440}}{f_{440}^{\rm GR}} - 1$, remains nearly flat, consistent with the prior.
In contrast, when the $(4, 4)$ multipoles are included, the posterior significantly deviates from the flat distribution (see Fig. 11 of ~\citet{LIGOScientific:2025obp}).
Applying the same analysis to GW150114 yields a similar behavior (see Fig. 4 of ~\citet{LIGOScientific:2025obp}), leading to the conclusion that a QNM on the hexadecapolar is likely present in GW250114.
This analysis further showed that the frequency of this \ac{QNM} agrees with the expected frequency of $440$ mode within tens of percent ($\delta \hat{f}_{440} = -0.06_{-0.35}^{+0.25}$).
The \texttt{KerrPostMerger} model~\cite{Gennari:2023gmx}, on the other hand, employs \ac{NR} calibrated amplitude for the post-peak waveform.
For GW250114, based on this waveform, the model including the $(4, 4)$ mode yields a $\log_{10}$ Bayes factor $0.54_{-0.18}^{+0.18}$ relative to the model including only the $(2, 2)$ mode, indicating that the data favor the presence of the $(4, 4)$ multipoles. 
Taken together, these results support the claim of the presence of the $440$ mode.

It is interesting to notice that the $440$ mode is not the only possible interpretation of the observed data.
Analyses of \ac{NR} simulations have shown that the quadratic mode $220 \times 220$ (for convenience, referred to as $220Q$) can be excited in the $(\ell, |m|) = (4, 4)$ radiation multipole with an amplitude comparable to that of the $440$ mode~\cite{Mitman:2022qdl, Cheung:2022rbm, London:2014cma, Cheung:2023vki, Gao:2025zvl, Nobili:2025ydt}.
This quadratic mode may produce an even higher \ac{SNR} in \ac{GW} detectors~\cite{Shi:2024ttu, Capuano:2025kkl}.
For GW250114, the \ac{SNR} ratio between $220Q$ and $440$ modes can be estimated based on the source parameters~\cite{Stein:2019qnm, London:2014cma, Cheung:2022rbm}, which is $\rho_{220Q} / \rho_{440} = 0.73/0.69 \approx 1$.
Moreover, the frequencies of the $220Q$ and $440$ modes are very close to each other~\cite{Mitman:2022qdl, Cheung:2022rbm}.
The frequency uncertainty of the $440$ mode inferred from GW250114 does not rule out the possibility that the observed mode actually contains the quadratic $220Q$ mode~\cite{LIGOScientific:2025obp}.
Therefore, the presence of the quadratic $220Q$ mode remains a highly plausible interpretation of the data.
In this Letter, we aim to use Bayesian inference to quantify the extent to which the data support the possibility that the quadratic $220 \times 220$ mode has been detected.


\textit{Ringdown Model.---}
Once the higher-order perturbations and nonlinear effects near the merger time have decayed, the ringdown waveform is dominated by a set of damped oscillations, known as \acp{QNM}.
The plus and cross polarizations of the \acp{QNM} in the detector can be expressed as
\begin{equation}
    h_{+} - \mathrm{i} h_{\times} = \sum_{\substack{\ell \geq 2 \\ 0 \leq m \leq \ell \\ n \geq 0}} A_{\ell mn} \mathrm{e}^{\mathrm{i} (2\pi f_{\ell mn}t + \phi_{\ell m n})} \mathrm{e}^{- t/\tau_{\ell mn}},
    \label{eq:ringdown_model}
\end{equation}
where $f_{\ell mn}$ and $\tau_{\ell mn}$ denote the Kerr frequencies and damping times, which can be solved from the Teukolsky equation \cite{teukolsky:1973perturbations} using first-order \ac{BH} perturbation theory, indexing by the angular-mode numbers $\ell$ and $m$, and the radial overtone number $n$.
The amplitudes $A_{\ell mn}$ and phases $\phi_{\ell mn}$ depend on the dynamic evolution of the progenitors.

Second-order \ac{BH} perturbation theory predicts the existence of second-order \acp{QNM}, which arise from the coupling of the first-order modes and are often referred to as \textit{quadratic \acp{QNM}}.
The parameters of quadratic modes are related to corresponding linear modes and can be written in the form~\cite{Mitman:2022qdl, Cheung:2022rbm, Nakano:2007cj}
\begin{equation}
    \begin{aligned}
        f_{\ell_{1}m_{1}n_{1} \times \ell_{2}m_{2}n_{2}} &= f_{\ell_{1}m_{1}n_{1}} + f_{\ell_{2}m_{2}n_{2}}, \\
        \tau_{\ell_{1}m_{1}n_{1} \times \ell_{2}m_{2}n_{2}}^{-1} &= \tau_{\ell_{1}m_{1}n_{1}}^{-1} + \tau_{\ell_{2}m_{2}n_{2}}^{-1}, \\
    \end{aligned}
\end{equation}
where $\ell_{1}m_{1}n_{1}$ and $\ell_{2}m_{2}n_{2}$ indicate pair of linear modes.

In this Letter, we perform a ringdown analysis with the Kerr template.
In this model, the frequencies and damping times of the \acp{QNM} depend on the mass and the spin of the remnant \ac{BH}, while the amplitudes are free variables.
The quadratic \acp{QNM} in the detector has the same form of signals as that of the linear modes, differing only in the way their frequencies and damping times are determined.
Therefore, the quadratic \acp{QNM} can be analyzed within the Kerr template in the same manner as ordinary \acp{QNM}.

\textit{Setup.---}
The GW250114 event was detected by the two LIGO detectors at 08:22:03 UTC on January 14, 2025.
Due to the very high \ac{SNR} (around 80), the signal was identified by multiple search pipelines \cite{KAGRA:2025oiz}.
Based on the \ac{IMR} analysis, it is determined that GW250114 is consistent with a merger of \ac{BBH}, corresponding to component masses of $m_1 = 33.6^{+1.2}_{-0.8} \, M_{\odot}$ and $m_2=32.2^{+0.8}_{-1.3} \, M_{\odot}$, with dimensionless spins constrained to $\chi_1 \leq 0.24$ and $\chi_2 \leq 0.26$.
The remnant \ac{BH} has a mass of $M_f = 62.7^{+1.0}_{-1.1} \, M_{\odot}$ and a spin of $\chi_f = 0.68^{+0.01}_{-0.01}$.
Furthermore, the analysis supports the presence of the $\ell = |m| = 4$ radiation multipole with a network \ac{SNR} of $3.6^{+1.4}_{-1.5}$.

Previous studies conclude that analysis results for ringdown signals are sensitive to the choice of the starting time.
Therefore, we perform Bayesian inference starting from the different times $t_0$, where the start time is characterized by $\Delta t = t_0 - t_{\rm peak}$ with $t_{\rm peak}=1420878141.235932$ GPS in geocenter.
Based on the redshifted mass of the remnant \ac{BH} $M_z = (1+z)M_f$, where $z$ denotes the cosmological redshift, we set the time interval between successive analyses to approximately $t_{M_z} = 0.337 \, \rm ms$.
The likelihood function is defined in the time domain, excluding all data before $t_0$.
In all analyses, we fix the sky location at $\alpha=2.333$rad, $\delta=0.190$rad, and the polarization angle at $\psi=1.329$rad \cite{KAGRA:2025oiz}.
Using the waveform templates that include different sets of \acp{QNM}, we sample over the the parameter space $\{M_z, \chi_f, A_{\mu}, \phi_{\mu}\}$, where $A_{\mu}$ and $\phi_{\mu}$ denote the amplitude and phase of each mode, respectively.
By comparing the parameter estimates and Bayesian evidence obtained from different models, we can identify which set of \acp{QNM} is more supported by the data.

We employ the \texttt{ringdown}~\cite{Isi:2021iql} software for the task of parameter estimation.
It employs the No-U-Turn Sampler~\cite{Hoffman:2011ukg} to obtain the samples to approximate the posterior distribution.
For nested models, such as the $220+221$ model and the $ 220$-only model, the Savage-Dickey ratio method~\cite{Dickey1971} can be used to estimate which model is better supported by the data.
Since some of the models we want to compare with are not nested, we extend the \texttt{ringdown} package to support sampling with \texttt{dynesty}~\cite{Speagle:2019ivv, sergey_koposov_2025_17268284}, thereby allowing for the estimation of the Bayesian evidence.
Both the Savage-Dickey ratio and the Bayesian evidence calculation can be used to construct the Bayes factor, which quantifies how the data prefer the explanation of one model over the competing model.

\begin{figure}[ht]%
    \centering
    \includegraphics[width=0.49\textwidth]{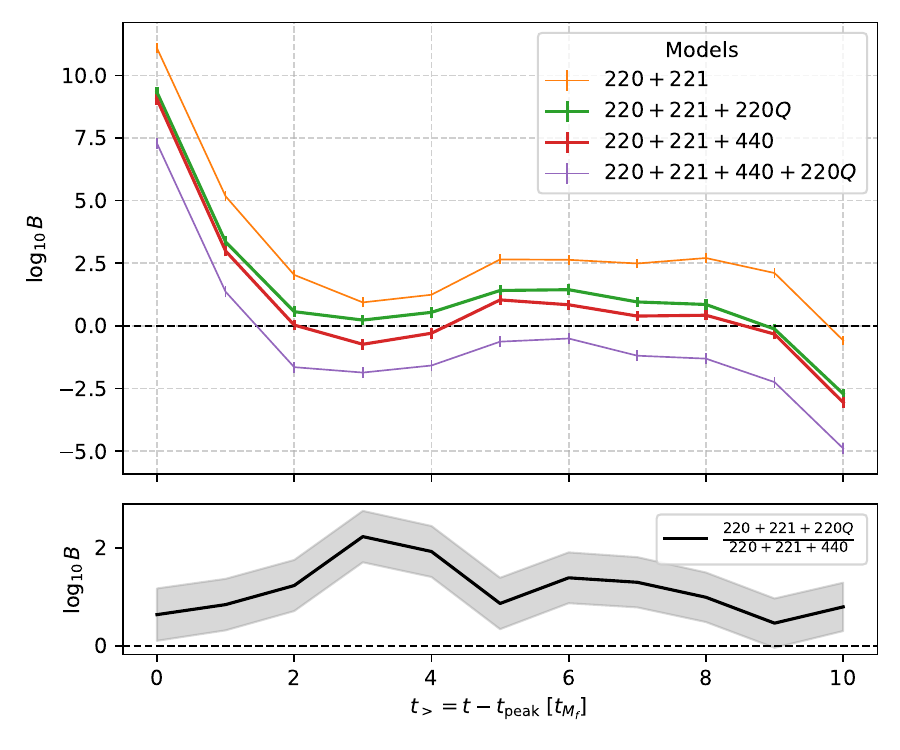}
    \caption{The upper panel shows the Bayes factors $\log_{10}B$ between the listed models against the $220$-only model, estimated at different start times. The lower panel presents the Bayes factor difference between the $220+221+220Q$ and $220+221+440$ models. $2\sigma$ error bars are given to indicate the estimation uncertainties. }
    \label{fig:bayes_factors}
\end{figure}

\textit{Result.---}
Using the analysis setup described above, we performed parameter estimation and model comparison at 11 different start times, from $t_{\rm peak}$ to $t_{\rm peak} + 10 t_{M_z}$, with a spacing of $1 t_{M_z}$ between successive analyses.
The Bayes factors $\log_{10}B$ between the target models against the $220$-only model are shown with error bars in Fig.~\ref{fig:bayes_factors}.

In this figure, the data favor the presence of only the $220+221$ modes.
The orange line is always the highest, no matter when the analysis begins.
On the contrary, the most complicated model, namely $220+221+440+220Q$, is consistently the least supported model.
In between are the models with three modes, where the third model can be the linear $440$ mode or the non-linear $220Q$ mode.
In most cases, the error bars are too small to overlap with other models.

For all four models, the Bayes factor generally drops as the starting time delays.
This can be explained by the fact that during the ringdown phase, the amplitude of the signal drops exponentially, so the less data analysed, the smaller the \ac{SNR} it contains, and it becomes harder to differentiate among different models.
For start times between $t_{\rm peak} + 3 t_{M_z}$ and $t_{\rm peak} + 5 t_{M_z}$, the Bayes factor all encounter an increase, this can be explained by the fact that the begining of the data is not purely \acp{QNM}, that the merger also can contribute a bit, and the noise in the data can also contribute to random fluctuation.

In comparing the results with three modes, we notice that the $220+221+220Q$ model is consistently more favoured compared with the $220+221+440$ model.
Although the two lines are quite close in Fig.~\ref{fig:bayes_factors}, in most cases the error bars do not overlap, indicating the data marginally support the existence of a nonlinear mode over the linear $440$ mode.
The exact numbers are not large enough to draw decisive conclusions on the actual composition of the $(4,4)$ mode; however, it is safe to say that the linear mode is not the only viable explanation, and with current observation, the possibility of a detection of the non-linear mode can not be ruled out.

\begin{figure}[ht]%
    \centering
    \includegraphics[width=0.48\textwidth]{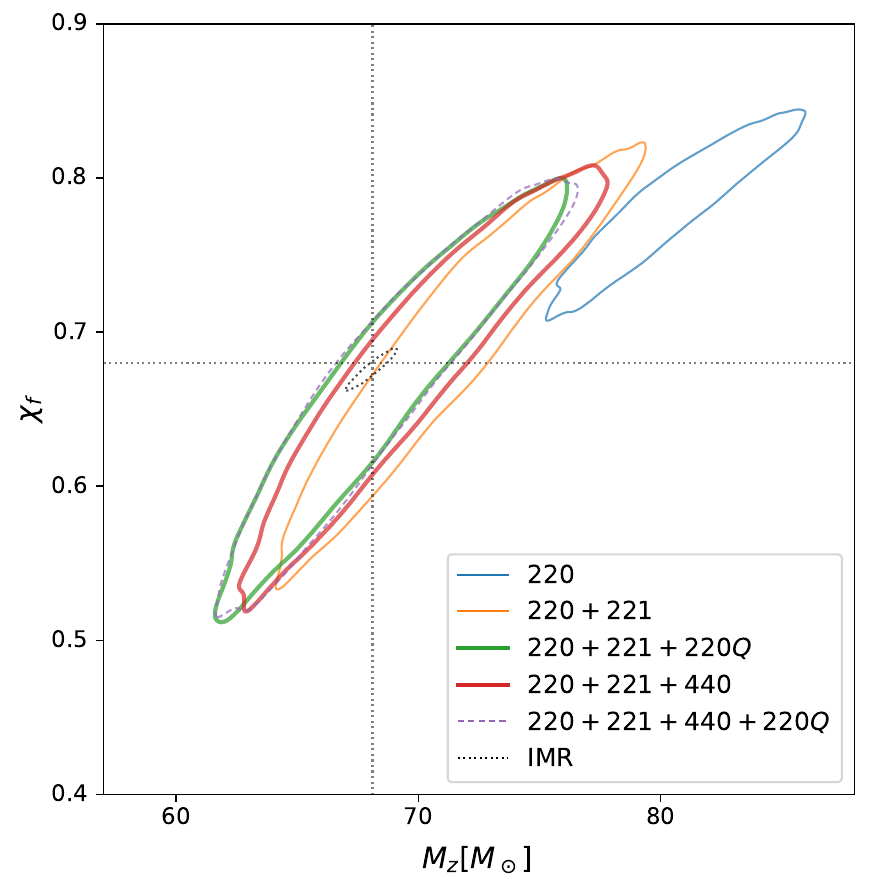}
    \caption{$90\%$ credible regions under different models for \ac{BH} mass and spin, using the data starting from $\Delta t = 4 t_{M_z}$.
   }
    \label{fig:mass_chi}
\end{figure}

Comparing against the $220+221$ model, the strongest support for the presence of $220Q$ mode occurs at $\Delta t=4t_{M_z}$.
Fig.~\ref{fig:mass_chi} shows the posterior estimates of the redshifted mass $M_z$ and spin $\chi_f$ of the remnant \ac{BH} at this start time for different models.
By comparing the 90\% credible regions, we find that only the ringdown models including three or more \acp{QNM} yield parameter estimates consistent with those from the \ac{IMR} analysis.
This result indicates that, with this start time, the $220+221$ model does not adequately describe the ringdown signal, whereas the $220+221+220Q$ or $220+221+440$ models provide a more self-consistent explanation under the \ac{GR} framework.
Furthermore, compared with the $220+221+440$ model, the $220+221+220Q$ model has a comparable size, but the IMR results sit in a more central region, indicating a better consistency. 

To quantify such consistency, we list the \ac{JSD}~\cite{Menendez1997} between the posterior from ringdown models and that from the \ac{IMR} model in Table~\ref{tab:JS_divergence}. 
A smaller value indicates a higher similarity between two distributions.
Due to the fact that the ringdown analysis excludes data that contains the majority of the \ac{SNR}, the 90\% credible regions are significantly larger than the IMR regions, and the \ac{JSD} is not small.
However, the relative values between different models are informative.
We find that the distributions obtained with $220Q$ mode agree more closely with that of the \ac{IMR} model than the ones including the $440$ mode, indicating that the ringdown data favor the presence of the $220Q$ mode.
We also report in Table \ref{tab:JS_divergence} the percentile rank of the maximum-likelihood point from the \ac{IMR} waveform within the posterior distributions obtained from different models.
This result further supports the above conclusion: for the models that do not include a $(4,4)$ mode, the maximum-likelihood estimate from the \ac{IMR} model was rejected at a 90\% confidence region.
While it locates around the 50\% credible region for the $220+221+440$ model, it fits the models with the non-linear mode the best, and locates within the 20\% credible region.

\begin{table}[ht]
    \centering
    \begin{tabular}{ccc}
        \hline
        Ringdown models & JSD & Percentile\\
        \hline
        $220$ & $0.693$ & $100.0\%$ \\
        $220+221$ & $0.672$ & $94.7\%$ \\
        $220+221+220Q$ & $0.579$ & $20.0\%$ \\
        $220+221+440$ & $0.618$ & $56.0\%$ \\
        $220+221+440+220Q$ & $0.576$ & $11.4\%$ \\
        \hline
    \end{tabular}
    \caption{Comparisons between different ringdown models against the \ac{IMR} model, allowing only the mass $M_z$ and spin $\chi_f$ to be variables. The middle column shows the JSD between posteriors, and the right column shows the percentile rank of the maximum-likelihood estimate from the IMR waveform under each model.}
    \label{tab:JS_divergence}
\end{table}

\textit{Conclusion and Discussion.---}
Based on our Bayesian analysis, we identify possible evidence for the contribution of the nonlinear $220Q$ mode in GW250114.
Although from a Bayesian model selection perspective, the data favor a model containing only two \acp{QNM}, $220$ and $221$, when considering a three \acp{QNM} model, the support for including $220Q$ is consistently stronger than that for $440$ mode.
Meanwhile, the posterior distribution of the frequency deviation supports the existence of the hexadecapolar mode.
Furthermore, a detailed examination at $\Delta=4t_{M_{z}}$ shows that the model including the $220Q$ mode yields a posterior more consistent with that of the full \ac{IMR} analysis.
The values of the Bayes factor are not large enough to draw decisive conclusions, but our analysis indicates that the possibility that a non-linear $220Q$ mode has already been observed can not be ruled out, as its power of explanation is as strong as the linear $440$ mode, if not stronger. 
Future analyses should therefore consider both these two modes simultaneously to distinguish which mode is responsible for the hexadecapolar radiation in the ringdown stage.

\begin{acknowledgments}

This work made use of and references the following software, listed in alphabetical order: \texttt{astropy} \citep{astropy:2013, astropy:2018, astropy:2022}, \texttt{dynesty}~\cite{Speagle:2019ivv, sergey_koposov_2025_17268284}, \texttt{Firefly-ringdown} \citep{Dong:2025igh}, \texttt{matplotlib} \citep{Hunter:2007}, \texttt{numpy} \citep{numpy}, \texttt{pandas} \citep{mckinney-proc-scipy-2010, pandas_17229934}, \texttt{python} \citep{python}, \texttt{Bilby} \citep{bilby_paper, bilby_paper_2, Bilby_17371955}, \texttt{corner.py} \citep{corner-Foreman-Mackey-2016, corner.py_14209694}, \texttt{Cython} \citep{cython:2011}, \texttt{h5py} \citep{collette_python_hdf5_2014, h5py_7560547}, \texttt{JAX} \citep{jax2018github}, \texttt{Numba} \citep{numba:2015, Numba_17206739}, \texttt{qnm} \citep{Stein:2019mop, qnm_3459790}, \texttt{ringdown}~\cite{Isi:2021iql}, \texttt{seaborn} \citep{Waskom2021}, and \texttt{tqdm} \citep{tqdm_14231923}.
Software citation information aggregated using \texttt{\href{https://www.tomwagg.com/software-citation-station/}{The Software Citation Station}} \citep{software-citation-station-paper, software-citation-station-zenodo}.

The authors want to thank Ziming Wang, Haitian Wang, Jiandong Zhang, and Jianwei Mei for constructive comments.
This work has been supported by the National Key Research and Development Program of China (No. 2020YFC2201400), the Natural Science Foundation of China (Grants No. 12173104), and the Fundamental Research Funds for the Central Universities, Sun Yat-sen University. 
\end{acknowledgments}

\section*{Supplemental material}

\begin{figure}[ht]%
    \centering
    \includegraphics[width=0.45\textwidth]{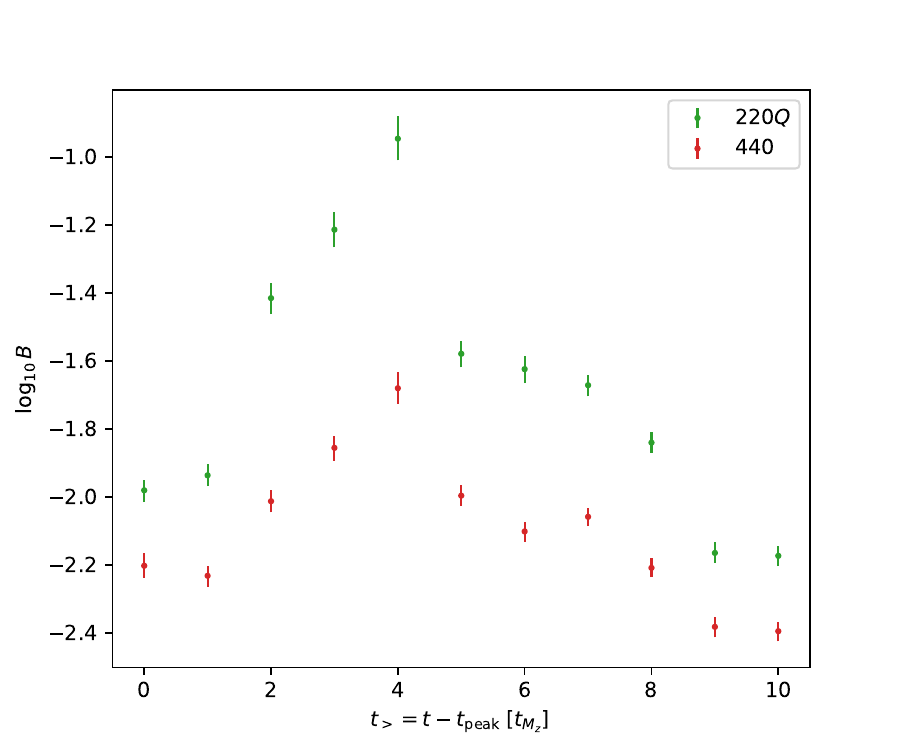}
    \caption{Bayes factors between the models that include $220Q$ mode/$440$ mode and those without, shown in green/red. Calculation obtained through the Savage-Dickey ratio method, with the error bars indicating $2\sigma$ credible intervals.}
    \label{fig:Savage_Dickey_Ratio}
\end{figure}

In Fig~\ref{fig:Savage_Dickey_Ratio}, we provide the Bayes factors estimated using the Savage-Dickey ratio method. 
The green/red lines show the values of logarithmic Bayes factors comparing models that include the $220Q$ mode/$440$ mode, in addition to the $220$ and $221$ modes, with those that do not, with different start times. The vertical error bars indicate the $2\sigma$ credible intervals. 
Although both cases provide little evidence in favor of the presence of additional modes, the level of support for the 440 mode is even lower

\bibliography{apssamp}

\end{document}